\documentclass[12pt]{iopart}

\usepackage{epsf}

\begin{document}

\newcommand{\bit}{\begin{itemize}}
\newcommand{\eit}{\end{itemize}}
\newcommand{\bc}{\begin{center}}
\newcommand{\ec}{\end{center}}
\newcommand{\be}{\begin{equation}}
\newcommand{\ee}{\end{equation}}
\newcommand{\beqn}{\begin{eqnarray}}
\newcommand{\eeqn}{\end{eqnarray}}
\newcommand{\ba}{\begin{array}}
\newcommand{\ea}{\end{array}}
\newcommand{\ra}{\rightarrow}
\newcommand{\lra}{\longrightarrow}

\title[Superconductor-to-Normal Phase Transition in Vortex Glass Model]
{Superconductor-to-Normal Phase Transition in a Vortex Glass Model:
Numerical Evidence for a New Percolation Universality Class}


\author{Frank O. Pfeiffer and Heiko Rieger}

\address{Theoretische Physik , Universit\"at des
Saarlandes, 66041 Saarbr\"ucken, Germany}


\begin{abstract}
  The three-dimensional strongly screened vortex-glass model is
  studied numerically using methods from combinatorial optimization.
  We focus on the effect of disorder strength on the ground state and
  found the existence of a disorder-driven normal-to-superconducting
  phase transition.  The transition turns out to be a geometrical
  phase transition with percolating vortex loops in the ground state
  configuration.  We determine the critical exponents and provide
  evidence for a new universality class of correlated percolation. 
\end{abstract}



\maketitle

\section{Introduction}

The gauge glass model is a paradigmatic model for disordered arrays of
Josephson junctions or amorphous granular superconductors
\cite{gaugeglass1}.  It has been argued that it also describes the
relevant physics of the superconductor-to-normal phase transition in
high-$T_c$ superconductors \cite{gaugeglass2}. A powerful tool to
investigate this transition is the domain wall renormalization group
(DWRG) technique that has been applied successfully to this model
\cite{dwrg,BY95,KS97,KR98}: In essence one calculates the stiffness of
the system with respect to twisting the phase variables at opposite
boundaries of a system of linear size $L$. If the twist costs an
energy that increases with $L$ one concludes that the system is
superconducting, if it decreases, one concludes that phase coherence
necessary for superconductivity is destroyed by thermal fluctuations,
i.e.\ the system is in a normal phase. In this paper we study this
model in the strong screening limit with varying strength of the
disorder at zero temperature. We will find a superconductor-to-normal
transition ({\it at $T=0$}) at a critical disorder strength and show
that it is accompanied by a proliferation of disorder induced global
vortex loop. By a finite-size scaling analysis of the loop statistics
we show that it is a percolation transition of a novel universality
class.

This paper is structured as follows: in section \ref{model} we present
the model a motivation to expect a disorder-driven phase transition
using the concept of the defect energy.
In the next both sections our results are presented. 
Section 3 shows a clear phase transition via the study of an
excitation loop perturbation.
In section 4 the transition is shown to be a geometrical phase
transition, what gives rise to apply percolation theory to the vortex
glass model.
The critical probability, above which a loop percolates, the critical
exponents and scaling relations are calculated numerically.
We close with a summary in section \ref{summary}.

\section{\label{model}Model}

The phenomenological lattice model describing the phase fluctuations
in a granular disordered superconductor close to the
normal-to-superconducting phase transition is the gauge glass model
\cite{BY95,KR98}
\be
{\cal H}  = - J \sum\limits_{\langle ij\rangle} \cos(\phi_i - \phi_j -
  A_{ij} - \lambda^{-1} a_{ij}) + \frac{1}{2}\sum\limits_{\opensquare} 
  (\nabla \times {\bf a} )^2,
\label{XY}
\ee
where $J$ is the effective coupling (set to 1) and $\phi_i$ the
phase on site $i$.
The sum is over all nearest neighbors $\langle ij\rangle$ on a simple
cubic lattice of system size $L$ with periodic
boundary conditions.
$A_{ij}$ are the vector potentials, which are uniformly distributed on 
\be
A_{ij} \in [0,2\pi \, \sigma] \qquad \mbox{with a fixed} \quad \sigma
\in [0,1],
\ee
where $\sigma$ defines the {\em disorder strength}.  $\sigma =1$
corresponds to strong disorder and $\sigma =0$ to the pure system,
respectively.  $\lambda$ is the bare screening length.  The
fluctuating vector potentials $a_{ij}$ are integrated over from
$-\infty$ to $\infty$ subject to ${\bf \nabla} \cdot {\bf a} = 0$.
The last term in (\ref{XY}) describes the magnetic energy and its sum
is over all elementary plaquettes of the lattice.  To investigate the
gauge glass model in the strong screening limit $\lambda \to 0$ we
make use of the vortex representation \cite{Kle89}, which gives after
standard manipulations \cite{BY95}
\be
\fl
{\cal H}_V^{\lambda \to 0} = 
\frac{1}{2}\sum_i ({\bf n}_i - {\bf b}_i)^2 \qquad
\label{H_V}
\mbox{with the magnetic field}
 \qquad {\bf b}_i = \frac{1}{2 \pi}\sum\limits_{\opensquare}A_{ij} 
\label{b}
\ee
subject to the local constraint $(\nabla\cdot {\bf n})_i = 0$. The
computation of the ground state of the Hamiltonian (\ref{b}), i.e.\ 
the vortex configuration ${\bf n}$ with the lowest energy ${\cal
  H}_V({\bf n})$, is a minimum-cost-flow problem that can be solved
{\it exactly} in polynomial time with appropriate combinatorial
optimization algorithms \cite{comb-review}.

We use the defect energy scaling method to show that there is a
superconducting-to-normal phase transition at low temperature $T$
varying the strength of disorder $\sigma$.
The idea is to calculate the energy $\Delta E$ necessary to introduce
a low-energy excitation loop (or domain wall) of size $L$ to the system.
We generate the excitation loop by a global manipulation of the energy
couplings along a fixed direction, as described in reference
\cite{KR98,PR99} in detail.
The defect energy $\Delta E$ results from the difference
energy of ground state with and without a global excitation loop.
Its disorder average is assumed to scale with the system size $L$ as
\be
\Delta E  \sim  L^{\theta}
\label{dE}
\ee
%
The sign of the stiffness exponent $\theta$ determines whether the
ground state is stable with respect to thermal fluctuations.  If
$\theta >0$ it costs an infinite amount of energy to induce a domain
wall crossing an infinite system ($L \rightarrow \infty$) and
therefore the the ground state remains stable at small but
non-vanishing temperatures: there is an ordered low temperature phase,
like in a 3$d$ XY-ferromagnet. On the other hand, if $\theta < 0$
arbitrarily large excitations loops cost less and less energy: the
ground state is unstable and thus not an ordered phase at any
non-vanishing temperatures, like in a 2$d$ XY spin glass.

We can easily see from (\ref{b}) that for small disorder (i.e.\ small
$\sigma$) the ground state is simply ${\bf n}={\bf 0}$ and $\theta=1$:
For a given disorder strength $\sigma$, it is $b_i \in [-2 \,\sigma ,
\;2 \,\sigma]$.  Thus, as long as $\sigma < 1/4$ it is $|b_i|<1/2$ and
the absolute minimum of all terms $({\bf n}_i - b_i)^2$ occurring in the
Hamiltonian (\ref{b}) fulfilling the constraint that ${\bf n}_i$ has to be
integer is ${\bf n}_i=0$.  A global excitation loop contains at least $L$
bonds with ${\bf n}_i=1$, which implies that for $\sigma < 1/4$ an
additional global excitation loop would cost a defect energy $\Delta
E\propto L$. This implies a stiffness exponent $\theta=1$ for small
disorder, certainly for $\sigma<1/4$, possibly even for larger sigma,
as we will see below. Thus we can already at this point conclude that
the for weak disorder the system described by (\ref{b}) is
superconducting (or ferromagnetic in the the magnetic, XY language),
as it is in the pure case ($\sigma=0$). 

On the other hand, in the opposite limit of strong disorder,
$\sigma=1$, defect energy calculations \cite{KR98,PR99} gave a
negative stiffness exponent $\theta= -0.96 \pm 0.05$, indicating the
absence of an ordered low-temperature phase (in particular the absence
of a stable low temperature vortex glass phase \cite{WY96}).
Therefore one can expect a disorder driven phase transition at zero
temperature from a superconducting phase for weak disorder to a normal
phase for strong disorder. We expect that this transition takes place
at a critical disorder strength $\sigma_c$ ($\sigma_c>1/4$ from what we
said above) and is characterized by a discontinuous jump of the
stiffness exponent $\theta$ from $1$ to $-0.96$ (here we assume the
simplest scenario in which one has only two attracting zero
temperature fixed points besides the critical point $\sigma_c$).

\begin{figure}[t]
\begin{center}
\epsfxsize9cm\epsfbox{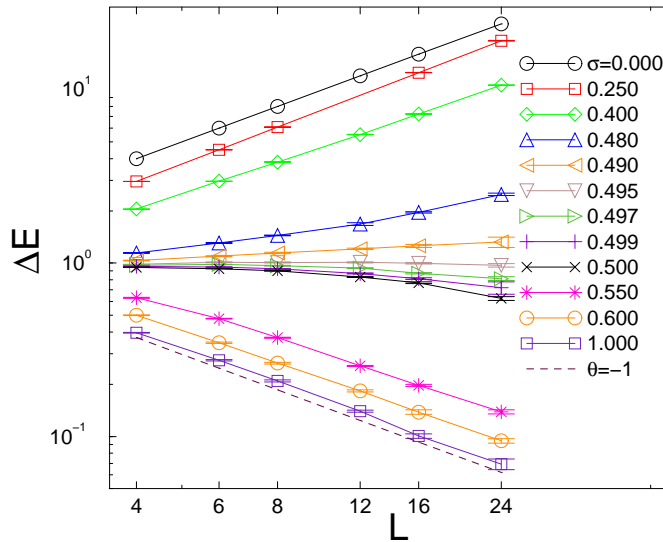} 
\caption{\label{H=0} \label{P_s} \label{p_s} \label{p_1L} 
  Log-log plot of the disorder averaged defect energy $\Delta E$ vs.
  system size $L$ for different disorder strengths $\sigma$.  For
  $L=4$, 6, 8 we used $N_{samp}=20000$ samples, for $L=12$ $\;5000$
  samples, $L=16$ $\;1000$ samples and $L=24$ $\;500$ samples,
  respectively.}
\end{center}
\end{figure}

\section{Defect Energy}

Figure \ref{H=0}, showing the defect energy $\Delta E$ versus
system size $L$ in a log-log plot, demonstrates that our numerical results
confirm this hypothesis. The slopes of the different curves,
representing different disorder strengths $\sigma$, are identical to
the stiffness exponent $\theta$. We observe that around $\sigma_c =
0.495 \pm 0.005$ it jumps from positive to negative with increasing
$\sigma$, this is our estimate for the location of the disorder driven
transition from the superconducting to the normal phase. Note that for
the unscreened gauge glass XY model it was found $\sigma_c \approx
0.55$ \cite{KS97}.

\begin{figure}
\epsfxsize16cm
\epsfbox{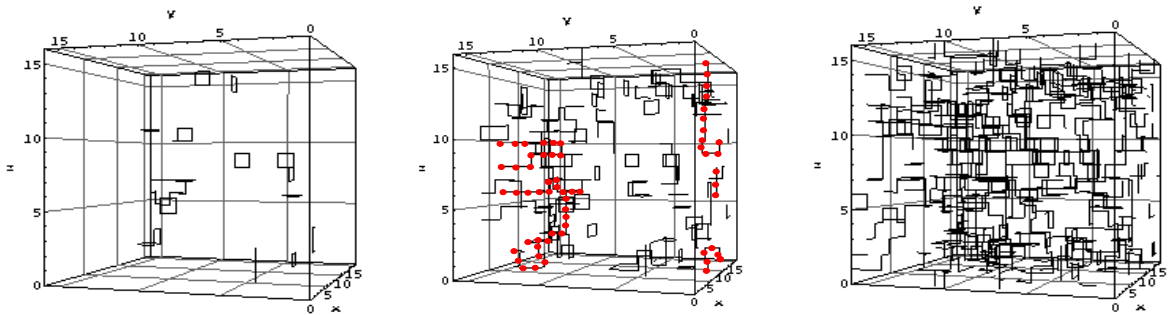}
\caption{Three ground state configurations for growing disorder
strengths $\sigma = 0.45$ (left), 0.50 (middle) and 0.55 (right), 
respectively, and system size $L=16$.
At $\sigma = 0.50$ it appears a percolating loop (dotted line).
}
\label{gs_conf}
\end{figure}

In what follows we will show that this zero temperature transition is
actually a 2nd order phase transition characterized by a single length
scale diverging at $\sigma_c$. This length scale corresponds to the
average diameter of closed loops in the ground state and at $\sigma_c$
these loops percolate the infinite system (note that we have periodic
boundary conditions), see Fig.\ \ref{gs_conf}. Thus what actually
happens at $\sigma_c$ is a percolation transition of vortex loops in the
ground state and in order to estimate its critical exponents 
we will perform a finite-size scaling analysis of the critical
behavior now.

\section{\label{expon}Vortex Loop Percolation Transition}

First we note that at $\sigma_c$ the concentration of vortex variables
that are non-zero (${\bf n}_i\ne0$), i.e.\ the probability $p$ with
which a bond in the simple cubic lattice is occupied with a vortex
segment, turns out to be $p_c=0.033\pm0.005$. This value is much
lower than the percolation threshold for conventional bond percolation
on the simple cubic lattice \cite{Sta85}, which is $p_c^{perco}
\approx 0.249$ \cite{Gra92}. This is a consequence of the global
constraint (divergence-free) underlying the optimization problem
(\ref{b}), which obviously causes strong correlations in the bond
occupation process.  Hence we suspect that the transition we are
considering establishes a new percolation universality class
\cite{remark}.

The geometrical objects of the ground state ${\bf n}$ of model
(\ref{b}) that we are going to study are {\it loops}. The algorithm to
detect loops is the following:\\ 
given the ground state configuration ${\bf n}$;\\
{\it while} it exists a vortex segment with ${\bf n}_i \not= 0$
along the bond $i$ do:\\
(I) choose $i$;\\
(II) find the shortest path
$P_i$ along non-zero vortex segments from the target site of $i$ to
the source site of $i$, where the direct path along $i$ is excluded;\\
(III) calculate $flow:= \sum_{j \in P_i \cup \{i\}} n_j$: if $flow
\not= 0$ or if there are two occupied bonds in a distance $L$ or
larger along the $x$, $y$ or $z$ direction, which belong to the same
loop, then the loop is called a {\it global} loop else a {\it local}
loop;\\
(IV) cancel the detected loop $P_i \cup \{i\}$ and continue the
{\it while} loop.

For each system with $L=4$, 6, 8, 10, 12, 16 we calculated 2000
different samples, for $L=20$, 24 and 32 1000 samples, respectively,
and then analyzed the loop statistics.

\begin{figure}
\begin{tabular}{cc}
\epsfxsize7.5cm
\epsfbox{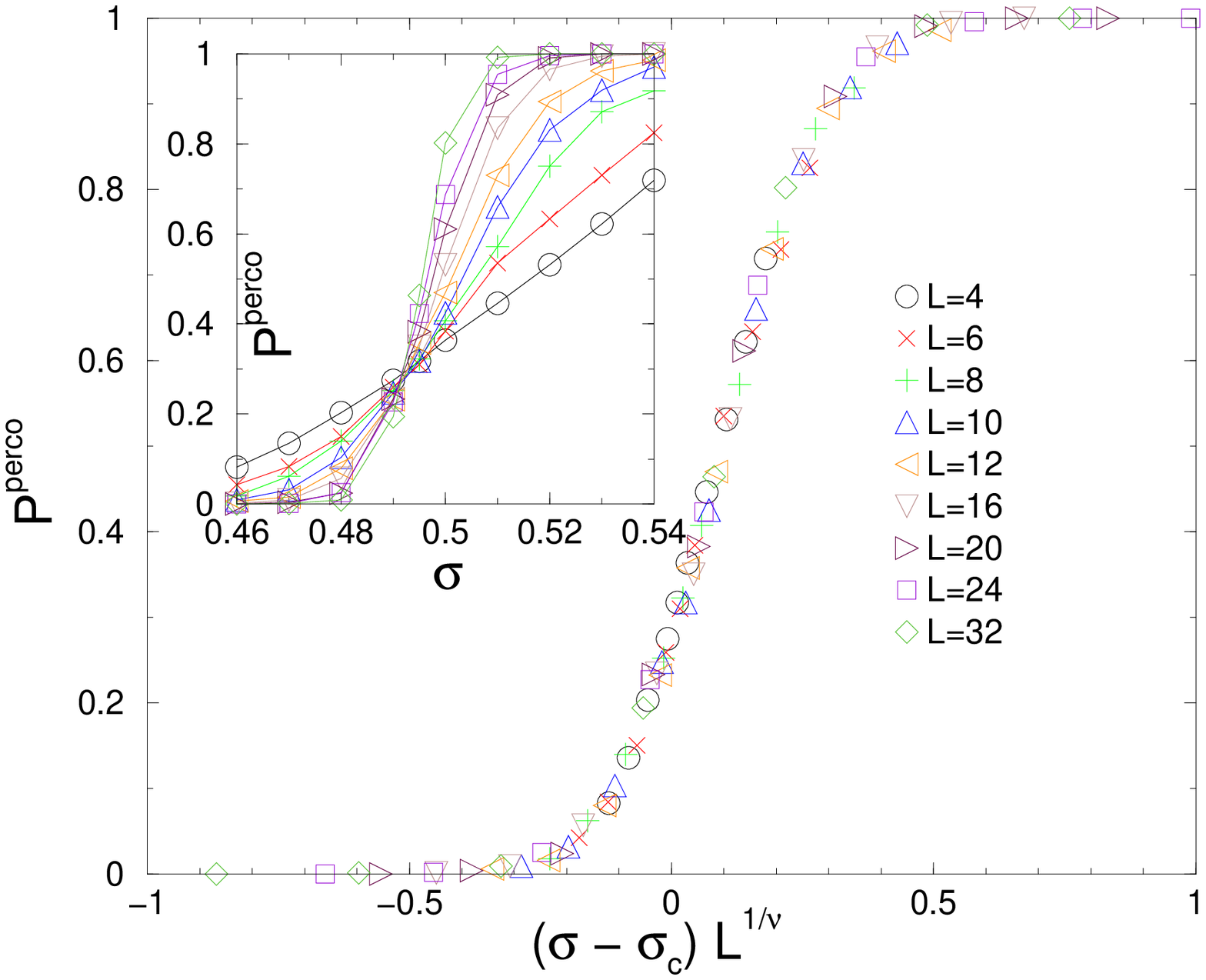} &
\epsfxsize7.5cm
\epsfbox{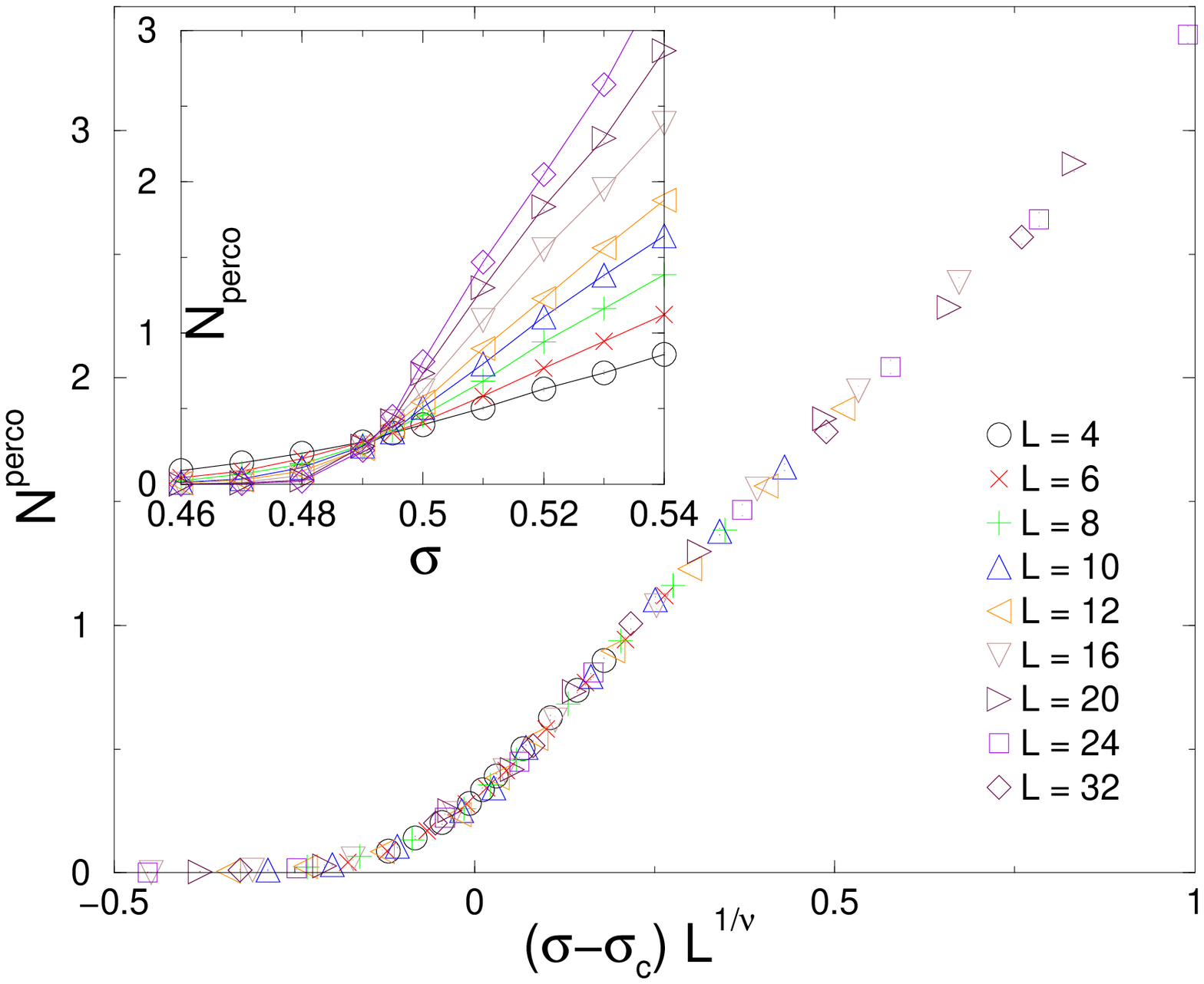}
\end{tabular}
\caption{\label{P_perco} \label{P_perco_scal} \label{N_perco}
 \label{N_perco_scal}
 Finite-size scaling of the percolation probability $P^{perco}$ (left)
 and the average number $N_{perco}$ of percolating loops (right) for
 different system sizes $L$ with $\sigma_c=0.492$ and $\nu=1.05$. The
 inset shows the raw data.  The error bars are smaller than the symbol
 size and therefore omitted.  }
\end{figure}

By studying the probability $P^{perco}_L(\sigma)$ that a system of
linear size $L$ contains at least one percolating loop we can check,
whether the percolation transition does indeed coincides with the
jump in the stiffness exponent located above. Its finite-size scaling
form is given by
\be
\fl
P^{perco}_L(p) = \tilde{P}^{perco}[\;(\sigma-\sigma_c) \cdot L^{1/\nu}\;]\;,
\ee
thus it is system size independent {\it at} $\sigma_c$ and curves for
different system sizes should intersect. Our data are shown in the
inset of Fig.~\ref{P_perco} (left) and we locate the intersection
point 
\be
\sigma_c=0.492 \pm 0.005
\ee
agreeing well with our estimate for $\sigma_c$ from the defect energy
analysis.

Next we deduce an estimate for the correlation length exponent $\nu$
by plotting $P^{perco}_L(p)$ versus $(\sigma-\sigma_c) \cdot
L^{1/\nu}$, where we fix $\sigma_c$ and determine $\nu$ such as
to achieve the best data collapse. This is done in Fig.~\ref{P_perco}
(left) and we obtain
\be
\nu=1.05\pm0.05.
\label{nu}
\ee
This estimate for $\nu$ lays between the value of the two- and
three-dimensional bond percolation \cite{Sta85}.

The analysis of the average number of percolating loops $N^{perco}_L(p)$,
obeying a similar finite-size scaling form $N^{perco}_L(p) =
\tilde{N}^{perco}[\;(\sigma-\sigma_c) \cdot L^{1/\nu}\;]$ gives the
same estimates for $\sigma_c$ and $\nu$, c.f.\ Fig.~\ref{P_perco}
(right). Note that {\it at } $\sigma_c$ the average number of
percolating loops does not (or only weakly) depend on the system size
and is small: $N^{perco}_L(p_c)\approx0.3$. The maximum number of
percolating loops we observed for $L=32$ at $\sigma_c$ was 3 with a
very low probability.

\begin{figure}
\begin{tabular}{cc}
\epsfxsize7.5cm
\epsfbox{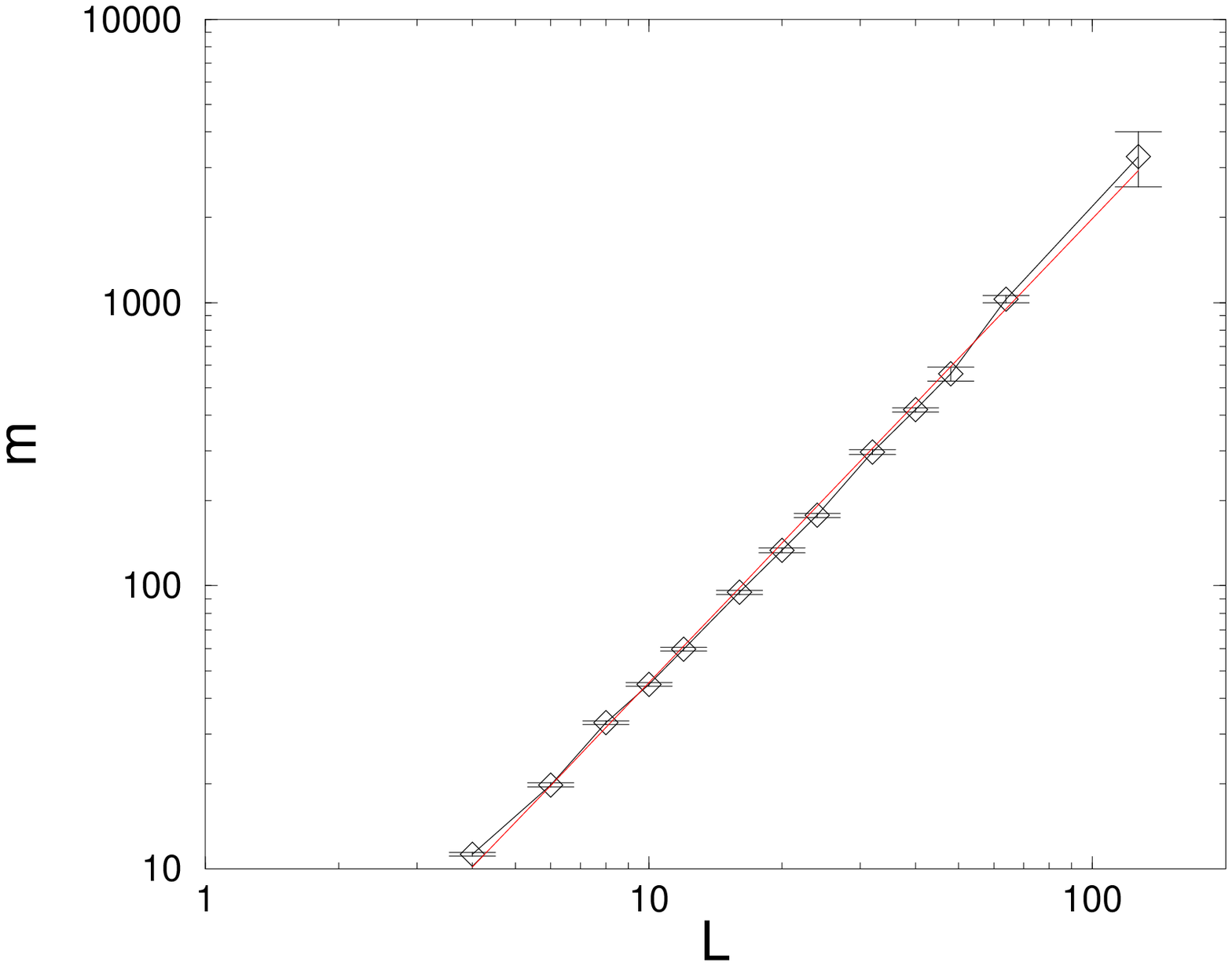} &
\epsfxsize7.5cm
\epsfbox{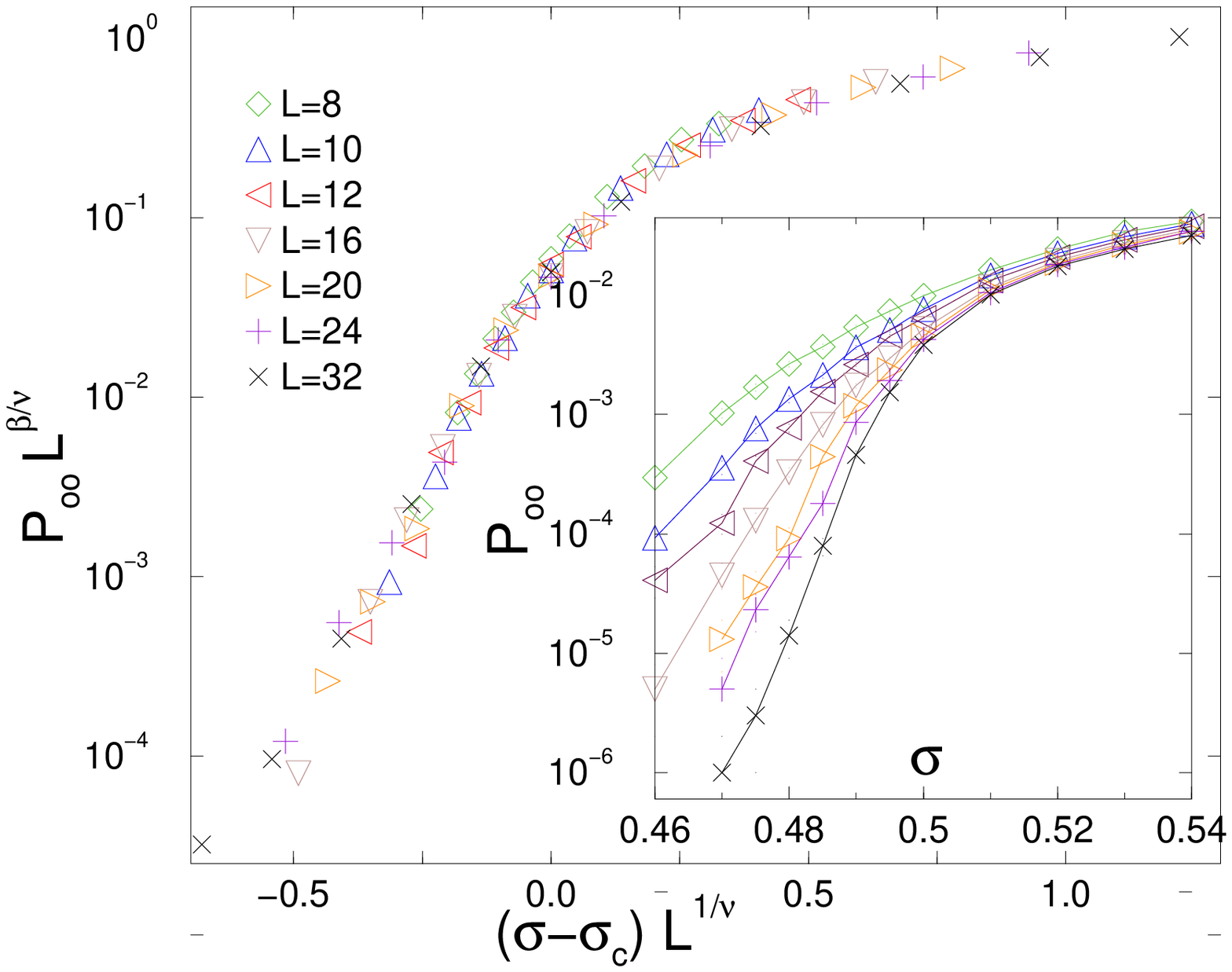}
\end{tabular}
\caption{\label{df_perco} \label{P_inf_s} \label{P_inf_s_scal}
  {\bf Left:} Plot of the average mass $m$ of a percolating loop vs.~$L$
  for $\sigma=\sigma_c$. The error bars are smaller than the symbols.
  The straight line is a least square fit to $m\sim L^{L^{d_f}}$ giving
  $d_f=1.64\pm0.02$ {\bf Right:} Finite-size scaling plot of the
  probability $P_{\infty}$ for a bond belonging to a percolating loop
  for different system size $L$ with $\sigma_c=0.495$, $\nu=1.05$ and
  $\beta/\nu=1.4$. The inset shows lin-log plot of the raw data.}
\end{figure}

The average mass $m$ of a percolating loop at $\sigma_c$ scales with 
$L$ like,
\be
m \sim L^{d_f},
\label{m-law}
\ee
where $d_f$ is a fractal dimension.
For $\sigma =\sigma_c$ we get with the data shown in 
figure \ref{df_perco}
\be
d_f=1.64 \pm 0.02\;.
\label{df}
\ee

The probability $P_{\infty}$ that a bond belongs to 
a percolating loop is expected to scale like 
\be
P_{\infty} \sim L^{-\beta / \nu} \; \tilde{P}_{\infty}[(\sigma
-\sigma_c) L^{1/\nu}].
\label{P_inf_scal}
\ee
The figure \ref{P_inf_s} (right) shows the raw data of $P_{\infty}$
(inset) and the plot of the scaling law (\ref{P_inf_scal}) with $\nu =
1.05 \pm 0.05$ and $\beta / \nu = 1.4 \pm 0.1$, i.e.
\be
\beta = 1.4 \pm 0.1.
\label{beta}
\ee
The usual hyper-scaling relation, $\beta/\nu=d-d_f$, known from
conventional percolation \cite{Sta85} gives $d_f = 1.6 \pm 0.1$, which is
consistent with (\ref{df}).

\begin{figure}
\begin{tabular}{cc}
\epsfxsize7.5cm
\epsfbox{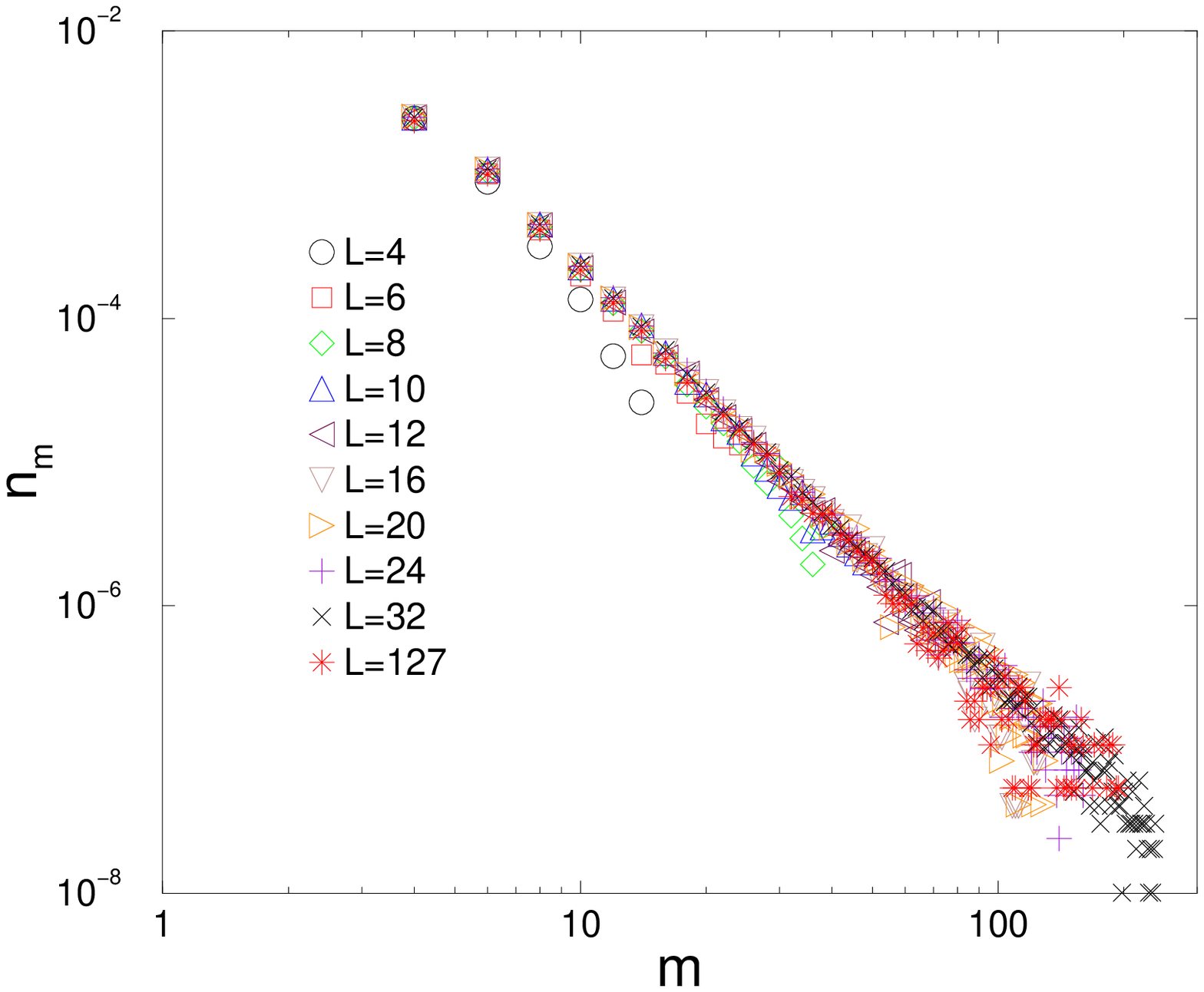} &
\epsfxsize7.5cm
\epsfbox{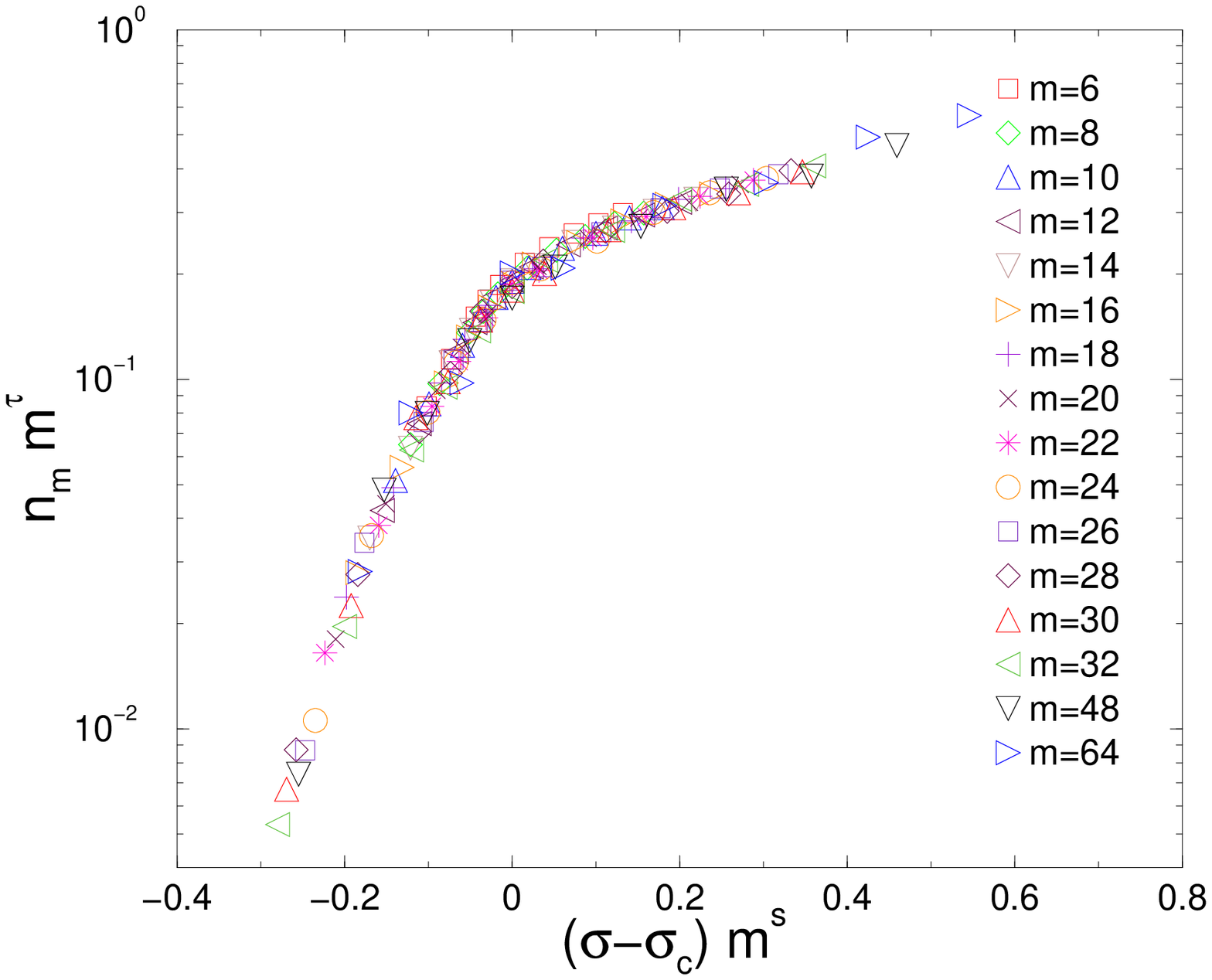}
\end{tabular}
\caption{\label{nm_m} \label{nm_s_scal}
  {\bf Left:} Probability distribution $n_m$ at $\sigma=\sigma_c$ for
  different system size $L$. A least square fit to $n_m\sim m^-\tau$
  yields $\tau=2.8\pm0.1$. {\bf Right:} Finite-size scaling of $n_m$ for
  $L=32$ with $\sigma_c=0.495$, $s=0.6$ and $\tau=2.95$.  For $m\ge30$
  the statistics is over less than 1000 loops for each $\sigma$.  }
\end{figure}

The loop distribution function $n_m$, i.e.\ the average number $n_m$
of finite loops of mass $m$ per lattice bond obeys the scaling form
(in the limit $L\to\infty$)
\be
n_m \sim m^{-\tau} \;\tilde{n}_m(\,(\sigma-\sigma_c)\,m^s),
\label{eq_nm}
\ee
where $\tau$ is the {\it Fisher} exponent and $s$ another critical
exponent (usually denoted $\sigma$ in conventional percolation, which
we avoid due to possible confusion with the disorder strength
$\sigma$).  The exponent $s$ describes how fast the number of loops of
mass $m$ decreases as function of $m$ close to $\sigma_c$.  Figure
\ref{nm_m} (left) shows the raw data of $n_m$ for different $L$ and
$\sigma=\sigma_c$.  For $L=32$ we get $\tau=2.89 \pm 0.05$ and for
$L=127$ (3 samples) $\tau=2.84 \pm 0.06$, respectively.  In the limit
$L \to \infty$ we expect
\be
\tau= 2.8 \pm 0.1.
\label{tau}
\ee
From the finite-size scaling plot of equation (\ref{eq_nm}), we get
$\tau=2.95 \pm 0.05$ and 
\be
s =0.6 \pm 0.1
\label{sigma}
\ee
for $\sigma_c  = 0.495\pm 0.005$ and $L=32$ in figure \ref{nm_s_scal} (right).

\begin{figure}
\begin{tabular}{cc}
\epsfxsize7.5cm
\epsfbox{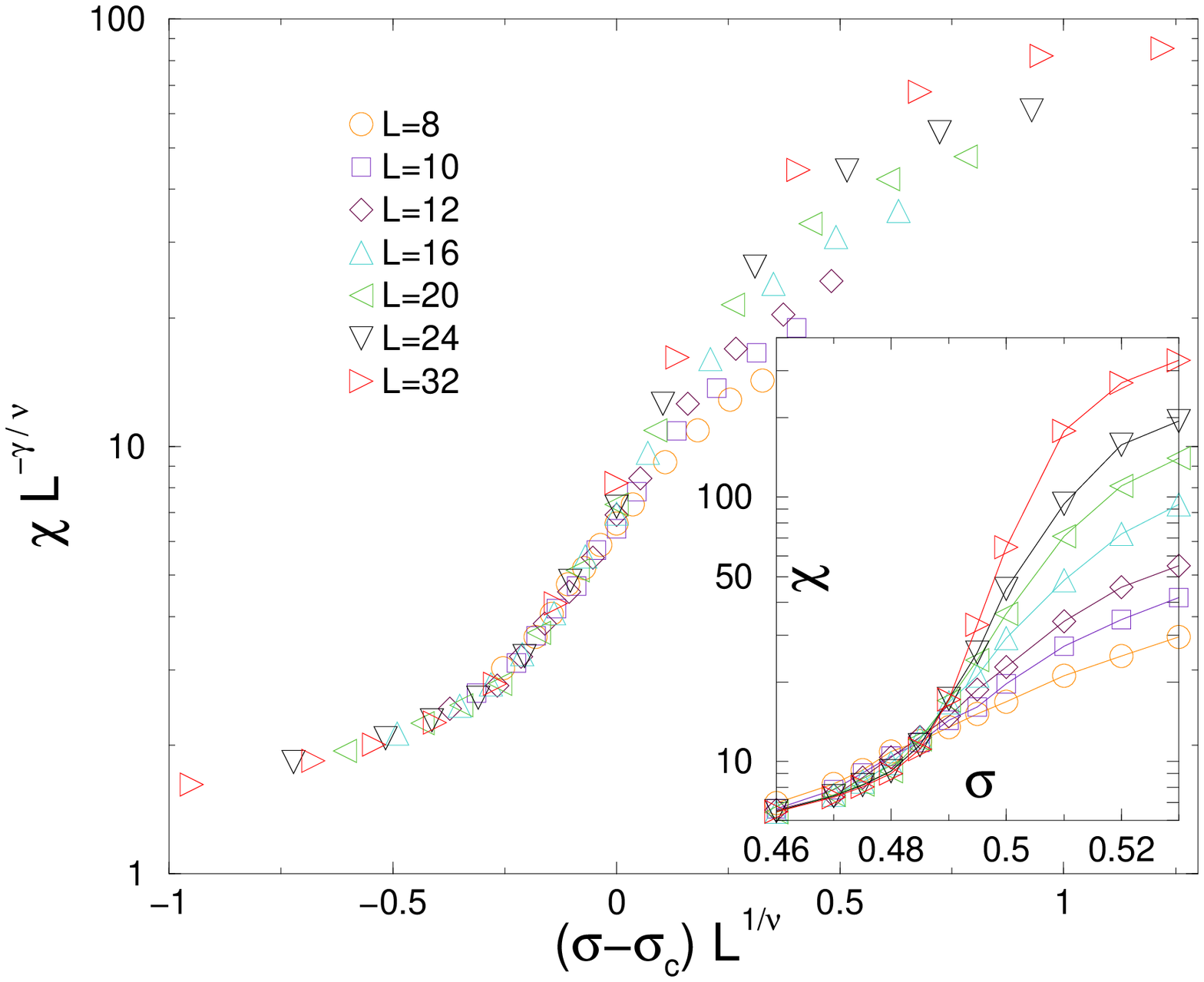} &
\epsfxsize7.5cm
\epsfbox{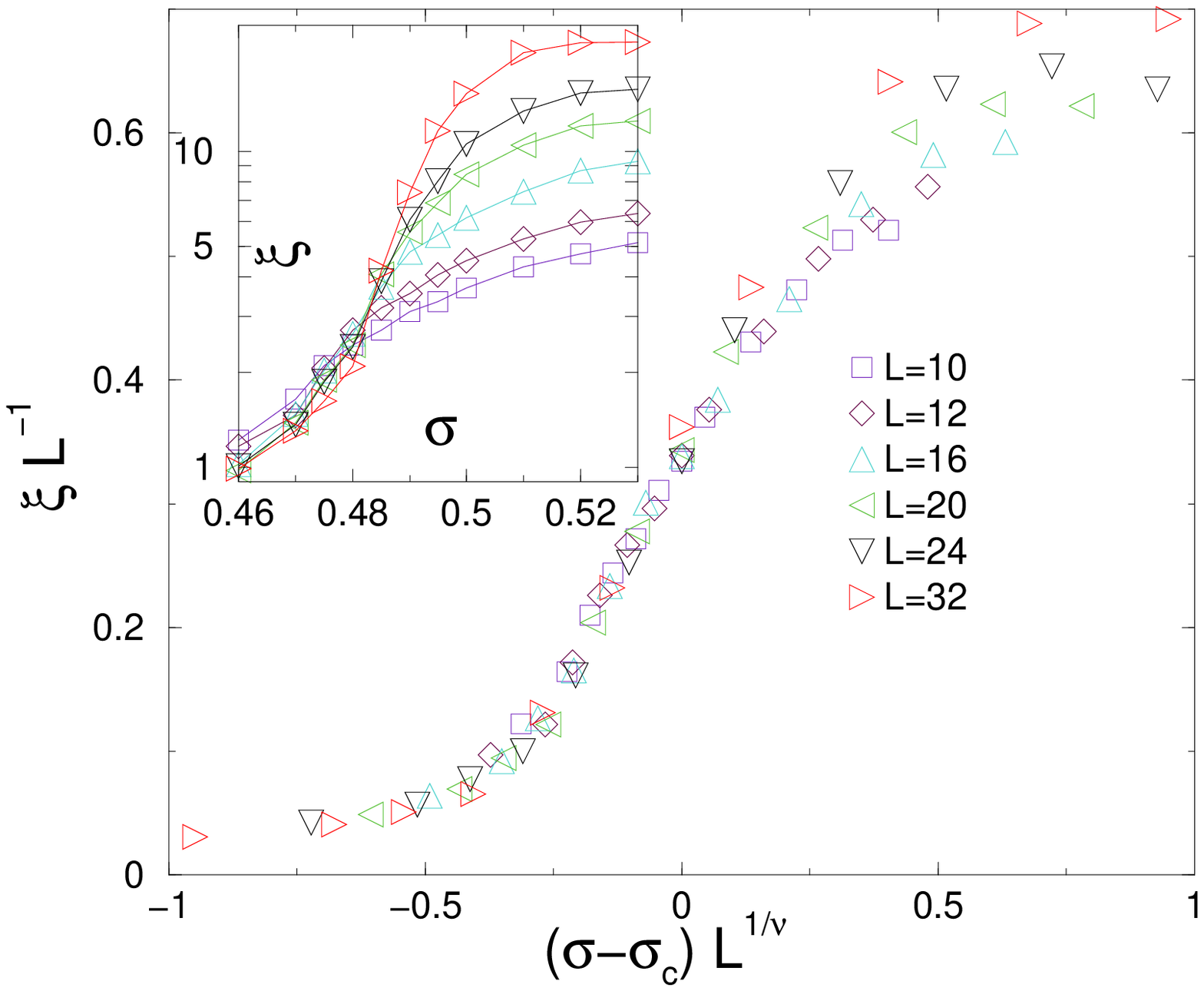}
\end{tabular}
\caption{\label{chi1} \label{chi_scal1} \label{xi} \label{xi_scal}
  {\bf Left:} Finite-size scaling plot of the susceptibility $\chi$ for
  $\sigma_c=0.495$, $\nu=1.05$ and $\gamma/\nu=0.42$ {\bf Right:} Scaling
  plot of the correlation length $\xi$ for $\sigma_c=0.495$ and
  $\nu=1.05$. The inset shows the raw data.}
\end{figure}

The zeroth moment $n = \sum_m n_m$ represents the average number of loops
per bond.  Below $\sigma_c$ the data collapse and satisfy $n \sim
\sigma$. The average loops size, defined as the ratio of the second
and first moment of the loop distribution \cite{Sta85}:
\be
\chi := \left(\sum_{m=4}^{\infty} m^2 n_m\right) /
\left(\sum_{m=4}^{\infty} m \, n_m\right).
\label{eq_chi}
\ee
is expected scale like 
\be
\chi \sim L^{\gamma/\nu} \; \tilde{\chi}[\;(\sigma-\sigma_c) \cdot
L^{1/\nu} \;].
\label{eq_chi_scal}
\ee
The data in figure \ref{chi_scal1} (left) show the raw data (inset) and
verify the scaling law (\ref{eq_chi_scal}) with $\gamma/\nu=0.4 \pm 0.1$
for $\sigma_c=0.495$ and $\nu=1.05 \pm 0.05$, i.e 
\be
\gamma=0.4 \pm 0.1.
\label{gamma}
\ee
The above estimates for $\gamma$ (\ref{gamma}) and $\beta$
(\ref{beta}) together with those for $\tau$ (\ref{tau}) and $s$
(\ref{sigma}) fulfill the usual exponent relation known from
conventional percolation
\be
\gamma = \frac{3-\tau}{s}\qquad,\quad
\beta= \frac{\tau-2}{s}\;.
\label{eq_gamma_scal}
\ee

Near $\sigma_c$ the linear size of a finite loop is characterized by
the correlation length $\xi$, which we calculate with the help of
the radius $R_{m_i}$ of gyration for the
loop $i$ of mass $m_i$ defined as
\be
R_{m_i}^2 := \frac{1}{m_i} \sum_{j=1}^{m_i} | {\bf r}_{ji}  - {\bf
r}_{0i}|^2 
\qquad
\mbox{with}
\qquad
{\bf r}_{0i} := \frac{1}{m_i} \sum_{j=1}^{m_i} {\bf r}_{ji}\;,
\ee
where ${\bf r}_{ji}$ is the position of a bond $j$ of the loop $i$ and 
${\bf r}_{0i}$ the center of mass, respectively.
Then, the correlation length $\xi$ is defined by
\be
\xi^2 := \left(\sum_{m=4}^{\infty} R_m^2 m^2 n_m\right) 
/ \left(\sum_{m=4}^{\infty} m^2 n_m\right)\;,
\ee
where $R_m$ is the average radius of gyration of loops of mass $m$
(averaged over disorder and individual loops). The raw data of $\xi$
are shown in inset of figure \ref{xi} (right). The finite-size scaling
form for $\xi$ is
\be
\xi \sim L \cdot \tilde{\xi}[\;(\sigma-\sigma_c) \cdot L^{1/\nu} \;].
\ee
From the best data collapse we get $\nu=1.05 \pm 0.05$, as shown in
figure \ref{xi_scal} (right), consistent with (\ref{nu}).

\begin{figure}
\begin{tabular}{cc}
\epsfxsize7cm
\epsfbox{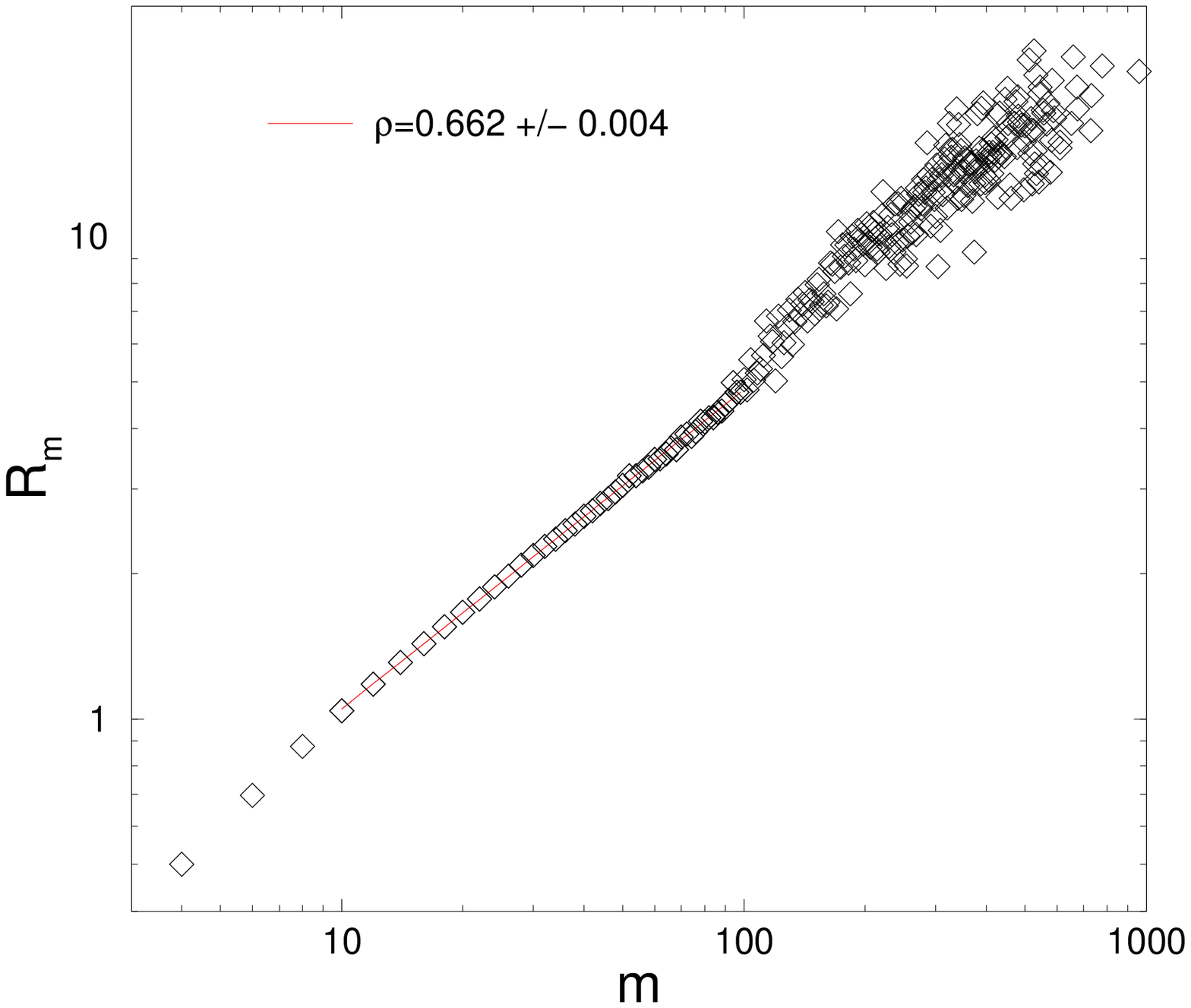} &
\epsfxsize7.5cm
\epsfbox{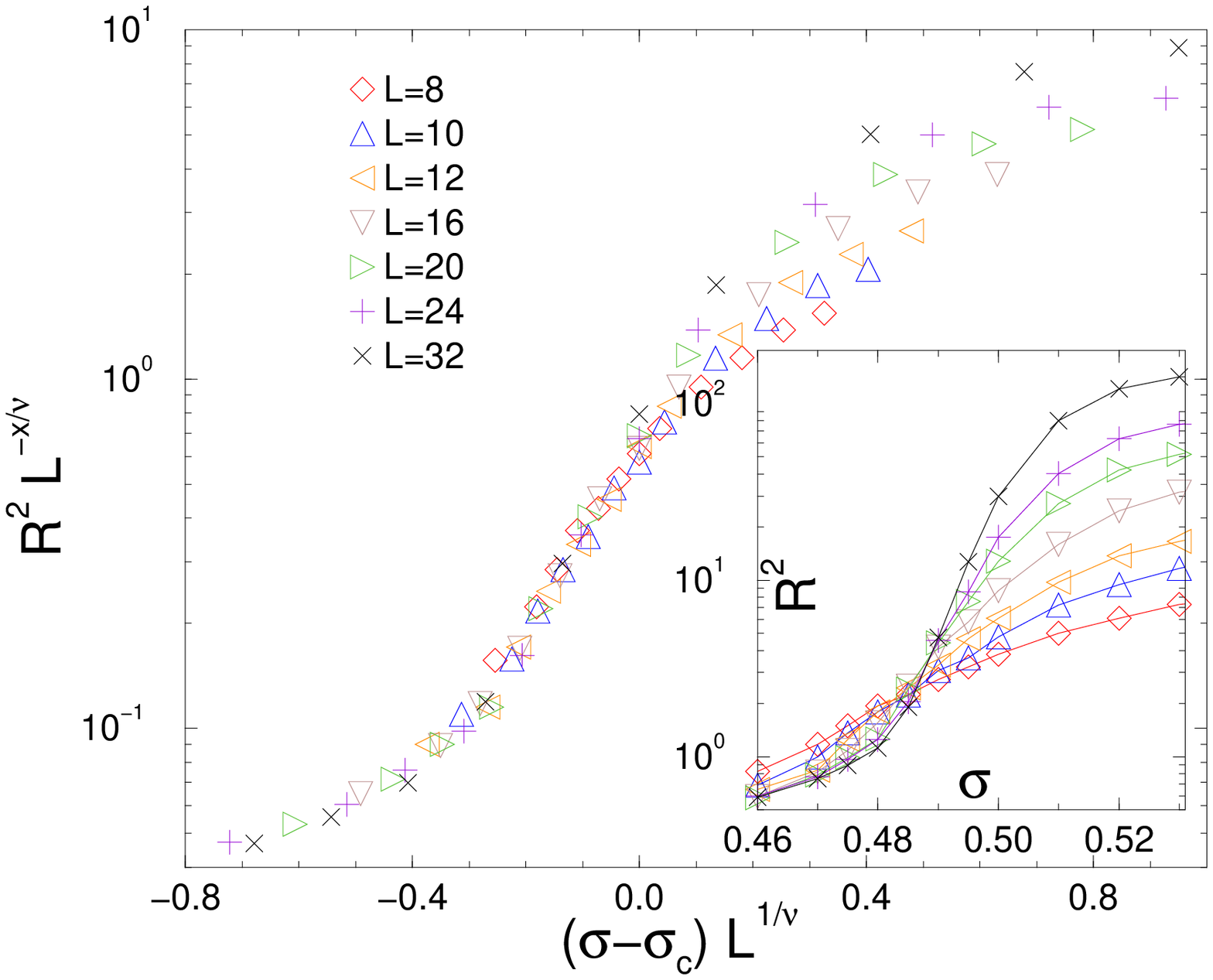}
\end{tabular}
\caption{\label{Rm_m} \label{R2_s}
  {\bf Left:} Average radius of gyration $R_m$ vs. the mass $m$ at
  $\sigma=\sigma_c$ for $1000$ samples and $L=32$. A least square fit
  to $R_m\sim m^\rho$ yields $\rho=0.66\pm0.02$. {\bf Right:} Finite-size
  scaling of the square loop radius of $R^2$ for different system
  sizes $L$ with $\sigma_c=0.492$, $\nu=1.05$ and $x/\nu=0.8$.  The
  inset shows the raw data.}
\end{figure}

At the percolation threshold the average radius of gyration $R_m$ of a
loop of mass $m$ increases algebraically
\be
R_m\sim m^\rho.
\label{rho-law}
\ee
In figure \ref{Rm_m} (left) we plot $R_m$ for $\sigma=\sigma_c$ and fit
the data in the interval $m \in \{10,...,100\}$ to the power law
(\ref{rho-law}), which yields
\be
\rho = 0.66 \pm 0.02\; \qquad \mbox{or} \qquad d_f=1/\rho=1.51 \pm 0.05\;,
\label{rho}
\ee
%
%
%
which agrees with our previous estimate for the fractal dimension $d_f$ 
of the percolating loops (\ref{df}) within the error bars.

Another quantity, which characterizes the size of finite loops, is the
mean square radius $R^2$, defined as
\be
R^2 := \left(\sum_{m=4}^{\infty} R_m^2 \; m \; n_m \right) 
/ \left(\sum_{m=4}^{\infty} m \; n_m\right).
\ee
We expect $R^2$ to scale like
\be
R^2 \sim L^{x/\nu} \; \tilde{R}[(\sigma -\sigma_c) L^{1/\nu} \;],
\ee
where $x$ is another critical exponent.  As depicted in figure
\ref{R2_s} (right) for the best data collapse we get $x/\nu = 0.8 \pm
0.1$ with $\nu=1.05 \pm 0.05$ (and $\sigma_c = 0.495 \pm 0.005$), i.e.\
\be
x=0.8\pm0.1.
\label{x}
\ee
This exponent should fulfill the relation \cite{Sta85}
\be
x=2\nu-\beta
\ee
With $\nu$ from (\ref{nu}) and $\beta$ from (\ref{beta}) we get
$x=0.7 \pm 0.2$, which is consistent with (\ref{x}).

\section{\label{summary}Summary}

In summary, we studied the ground state of the three-dimensional
strongly screened vortex glass model, numerically.  We found a clear
evidence for a disorder-driven superconducting-to-normal phase
transition indicated by a change in the stiffness exponent at
$\sigma_c$. This transition turned out to be a percolation transition
for disorder induced vortex loops crossing the whole system. 

At first sight it might be surprising, why the existence of percolating
vortex loops is related to a change in the stiffness exponent of model
(\ref{XY}).  However, the the stiffness exponent provides information
on how hard it is to induce a domain wall into a system of linear size
$L$ and a domain wall is surrounded by a global vortex loop. If, at
and above a critical disorder strength, global vortex loops
proliferate already in the ground state, the creation of an extra
excitation loop will, with probability one, costs only an infinitesimal
amount of energy in the infinite system size limit.

A similar observation --- the coincidence of vortex loop percolation
and a thermal phase transition in superconductors --- has been made
earlier in models for high-$T_c$ superconductors: In \cite{NS98} it
was shown for a model of a {\it pure} superconductor that the melting
transition of the Abrikosov flux line lattice at the temperature
$T_{c2}$, where the transition from the superconductor to normal phase
takes place, is accompanied by a proliferation of thermally induced
global vortex loops. And similarly in \cite{JB96} it was shown that
the temperature driven resistivity transition in disordered high-$T_c$
superconductors is also accompanied by a percolation transition of
vortex lines perpendicular to the applied field. These thermally
induced transitions are, however, in universality classes different
from the disorder induced transition we studied here.

\ack 
We thank M.\ Kosterlitz for pushing us to study the model
(\ref{b}) with varying disorder strength and for stimulating
discussions. This work was supported by the Deutsche 
Forschungsgemeinschaft (DFG).


\begin{thebibliography}{<num>}
\bibitem{remark} A different loop percolation transition happens in
  plaquette percolation, in which randomly, with probabiliy $p$, the
  elementary plaquettes of a simple cubic lattice are occuppied, the
  4 boundary-sides forming an elemantry loop. Here the optimization
  constraint coming from the Hamiltonian (\ref{b}) is missing, giving
  rise to a different universality class than the one considered here.
\end{thebibliography}
\end{document}